\documentclass[fleqn,usenatbib]{mnras}

\usepackage{newtxtext,newtxmath}

\usepackage[T1]{fontenc}

\DeclareRobustCommand{\VAN}[3]{#2}
\let\VANthebibliography\thebibliography
\def\thebibliography{\DeclareRobustCommand{\VAN}[3]{##3}\VANthebibliography}

\usepackage{graphicx}	
\usepackage{amsmath}	
\usepackage{soulutf8}
\usepackage{booktabs,tabularx}
\usepackage[normalem]{ulem}
\usepackage{float}

\newcommand{\nfrac}{\genfrac{}{}{0pt}{}}

\title[]{Constraining mass, radius and tidal deformability of compact stars with axial $wI$ modes: new universal relations including slow stable hybrid stars}     

\author[Ignacio F. Ranea-Sandoval et al.]{
Ignacio F. Ranea-Sandoval$^{1, 2}$\thanks{iranea@fcaglp.unlp.edu.ar},
Mauro Mariani$^{1, 2, 3}$,
Germ\'an Lugones$^{3}$, and
Octavio M. Guilera$^{1, 4}$
\\
$^1$ Grupo de Gravitaci\'on, Astrof\'isica y Cosmolog\'ia, Facultad de Ciencias Astron{\'o}micas y Geof{\'i}sicas, Universidad Nacional de La Plata, \\Paseo del Bosque S/N, 1900, La Plata, Argentina.\\
$^2$ CONICET, Godoy Cruz 2290, 1425, CABA, Argentina.\\
$^3$ Centro de Ciências Naturais e Humanas, Universidade Federal do ABC,  Avenida dos Estados 5001, CEP 09210-580, Santo André, SP, Brazil.\\
$^{4}$ Instituto de Astrofísica de La Plata, CCT-La Plata, CONICET, Paseo del Bosque S/N, 1900, La Plata, Argentina.
}

\date{Accepted 2022 December 19. Received 2022 December 19; in original form 2022 October 11}

\pubyear{2022}

\begin{document}
\label{firstpage}
\pagerange{\pageref{firstpage}--\pageref{lastpage}}
\maketitle

\begin{abstract}
We revisit asteroseismology with quadrupolar $wI$ modes and present universal relationships for its fundamental and first overtone. In contrast to relationships proposed in the literature, our universal relationships are capable of including slow stable hybrid stars that appear when considering slow sharp hadron-quark phase transitions. We show that, if the frequency and damping time of the fundamental mode of a given pulsating object are measured, its mass, radius, and dimensionless tidal deformability can be inferred. Moreover, we show that the errors of such estimates are smaller than a few percent for the mass and radius. For the dimensionless tidal deformability, the error{s are -for compact objects with $M\gtrsim 1.4\,M_\odot$- in general smaller than $\sim 100 \,\%$}. Comparison with previous universal relationships shows that the ones proposed in this work produce better estimates of the mass and radius of totally stable compact objects.
\end{abstract}




\section{Introduction}\label{sec:intro}

{Neutron stars (NSs) can emit gravitational waves (GWs) by \mbox{space-time} pulsational modes generically called axial $w$ modes with typical frequencies in the range of $\sim 5-20$~kHz. Contrary to the widely studied polar fluid modes \citep[see the modern review by][and references therein]{andersson2021Univ}, axial $w$ modes barely excite any fluid motion and are rapidly damped in $\sim 10^{-4}$~s. However, it has been suggested that they can be excited in the collapse of a NS into a black hole (BH) soon before the BH formation \citep{Baiotti:2005vi}, as well as by a mass scattered by the NS  \citep{Ferrari:1999gp,Andrade:1999mj}.   Although $w$ modes frequencies are above the range of high sensitivity of current gravitational wave observatories, future third generation interferometric arrays such as the Einstein Telescope and the Cosmic Explorer will improve significantly the chances of detection \citep{maggiore:2020scf,hall:2022cea}. It is therefore important to develop and refine theoretical tools to extract the astrophysical information carried by these modes.  }

In \citet{RSetal2022PRD}, we have shown that the two first quadrupolar ($\ell =2$) axial $wI$ modes of \emph{slow stable hybrid stars} (SSHS) do not follow any of the \textit{universal relationships} (URs) proposed in the literature \citep[see, for example,][]{Benhar2004PRD,Tsui2005MNRAS,w-modes_universal}.  SSHSs are hybrid stars (HSs) that are stable even if the derivative of the mass with respect to the central energy density is negative \citep{Pereira:2017rmp,LugGrunf-universe:2021,lugones2021arXiv,RSetal2022PRD}. This is possible if there is an abrupt hadron-quark interface in the stellar interior and if the conversion speed at the phase-splitting surface is \emph{slow} when the star is perturbed. This scenario gives rise to an extended twin branch of SSHSs in the mass-radius plane. {We emphasise that several SSHSs of the extended branch have central energy densities above thirty times the nuclear saturation one. On the contrary, if the conversion is \emph{rapid}, the stability remains while the derivative of the mass with respect to the central energy density is positive and up to the maximum mass star.} In principle, SSHSs are possible objects of the Universe as they are stable against radial perturbations in the same sense as any compact object \citep{Pereira:2017rmp,LugGrunf-universe:2021,lugones2021arXiv}. This fact put restrictions on the applicability of current URs to determine astrophysical quantities of compact objects.

In the new era of multimessenger astronomy with GWs, the relevance of asteroseismology of compact objects can not be understated. This branch of compact object's astrophysics relies on the fact that quantities related to the frequency and damping time of particular quasi-normal modes (QNMs) correlate with macroscopic quantities associated with the pulsating object, (almost) independently of the description of matter inside the compact object.  
Asteroseismology of NSs started, from a theoretical point of view, after the seminal paper by \citet{AK} (for more details see the reviews by \citet{ferrari-gualtieri2008QNMs,schutz2008ans} and references therein). Following this paper, several improvements have been proposed and different approaches explored (see, for example, \citet{chirnti-f-mode2015,IloveQw2021,sotanikumar2021}, and references therein).

\begin{figure*}
\centering
\includegraphics[width=0.7\linewidth,angle=0]{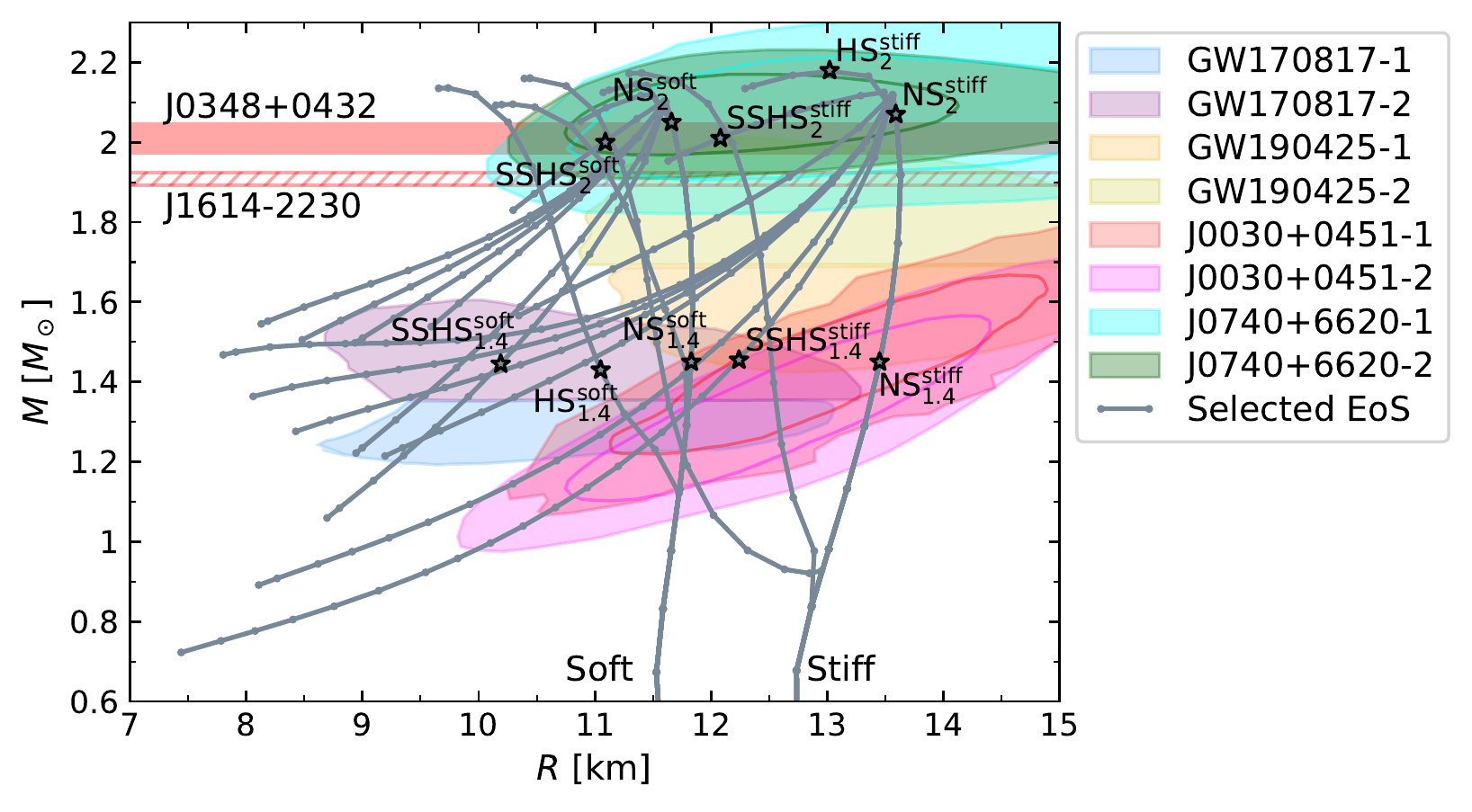}
\caption{Mass-Radius relationships obtained with the hybrid {EoSs} used in this paper. With colours, different modern astronomical observations of compact objects. The ten synthetic objects for which the proposed URs have been applied are also indicated with black symbols (for details regarding the names, see Table~\ref{tabla:application} and Section~\ref{sec:astro}).}
\label{fig:MR}
\end{figure*}

In this paper, we present new URs for the fundamental axial $wI$ mode and the first overtone that are valid not only for hadronic NSs and fully stable HSs (i.e. HSs that are stable for any reaction speed at the interface), but also for SSHSs. {We will only focus our attention on URs with potential astroseismological applicability to estimate macroscopic parameters (such as the gravitational mass, $M$, the radius, $R$, and the dimensionless tidal deformability, $\Lambda$) of the pulsating object.} In addition, we show how URs for the fundamental mode might be useful, in the future, to estimate mass, $M$, radius, $R$, and dimensionless tidal deformability, $\Lambda$, of a pulsating object if the frequency, $\nu$, and damping time, $\tau$, of such mode can be detected. Moreover, we compare our fits to the one presented in \citet{Benhar2004PRD}; we only compare with this work because it is the only one for which we have found errors in the fitting parameters of the URs for axial $wI$ modes, giving us the possibility to compare not only the values predicted by the asteroseismology method but also the associated errors bands. We show that the estimates of mass and radius obtained with our proposed relationships are finer even for purely hadronic NSs. In addition, we show that $wI$ mode asteroseismology might be able to distinguish between NSs and SSHS twins.

The structure of this paper is the following. In Section \ref{sec:newUR}, we present the proposed URs for axial $wI$ modes. In Section \ref{sec:astro}, we present the astronomical capabilities of the proposed URs and compare them with those of previous URs. A summary of the work and a discussion about the astrophysical implications of our results are provided in Section \ref{sec:conc}. In Appendix \ref{app:fits}, we present the tables with the coefficients and additional aspects of the URs proposed for the first overtone.

\begin{figure*}
\centering
\includegraphics[width=0.475\linewidth,angle=0]{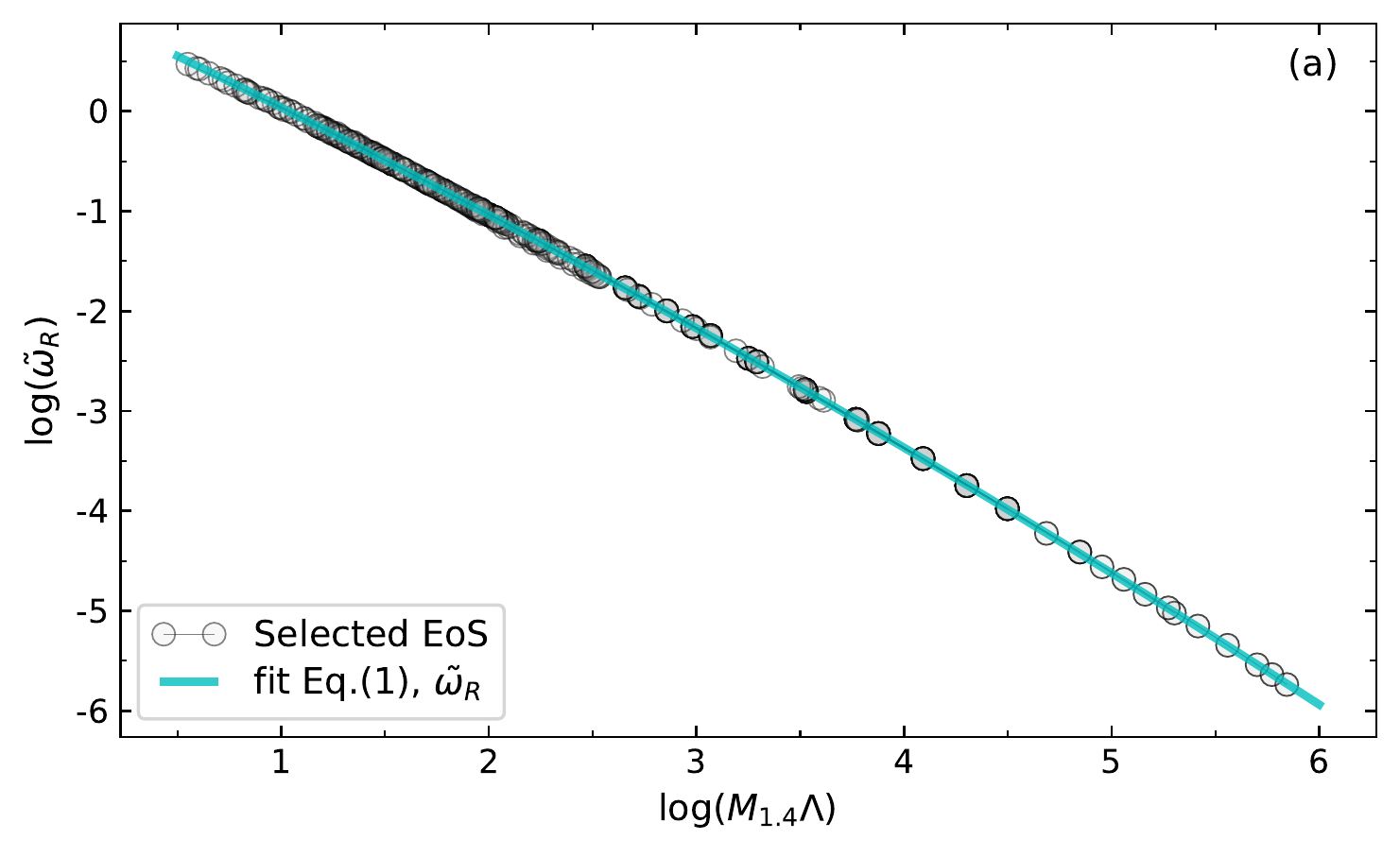}
\includegraphics[width=0.475\linewidth,angle=0]{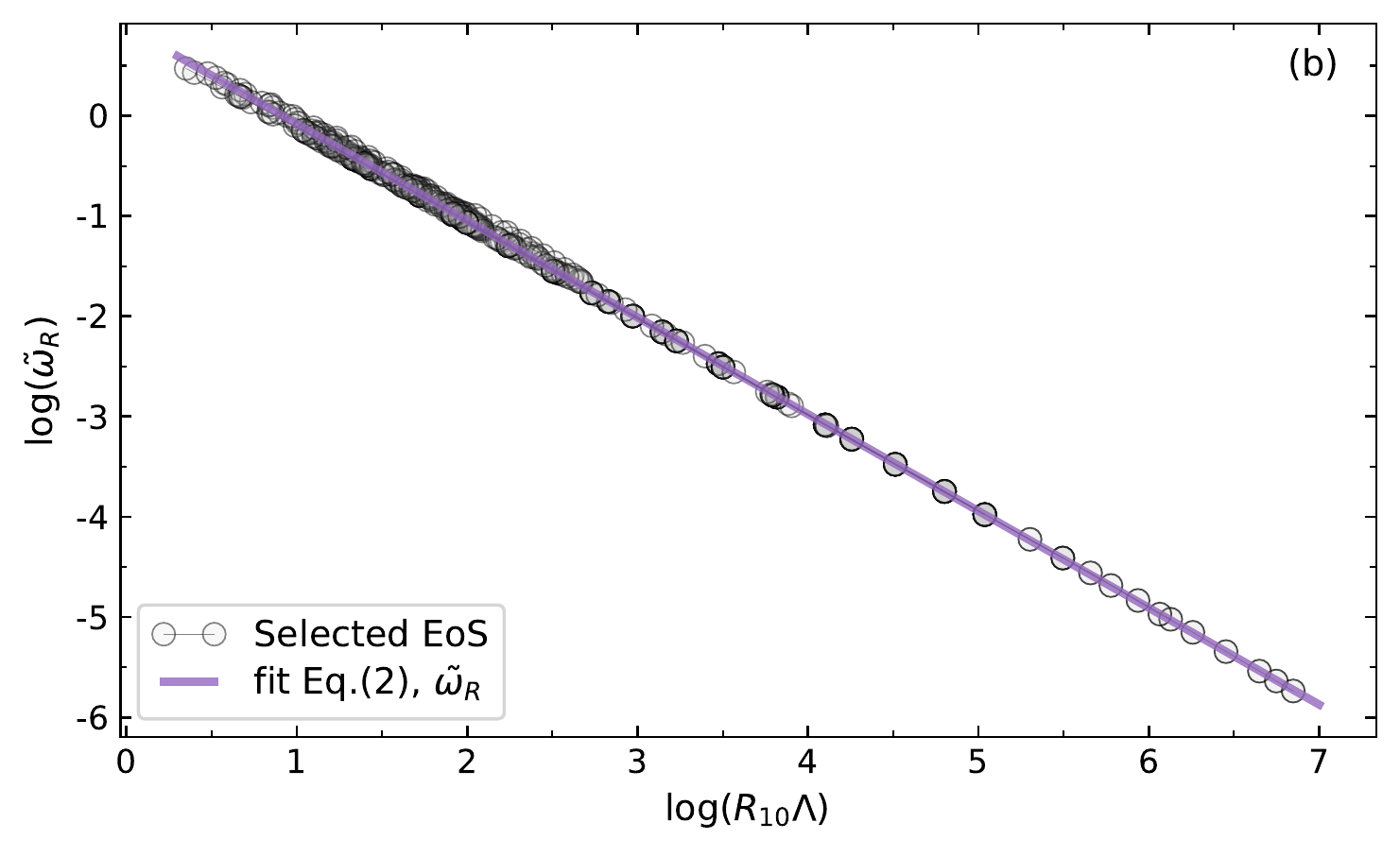}
\includegraphics[width=0.475\linewidth,angle=0]{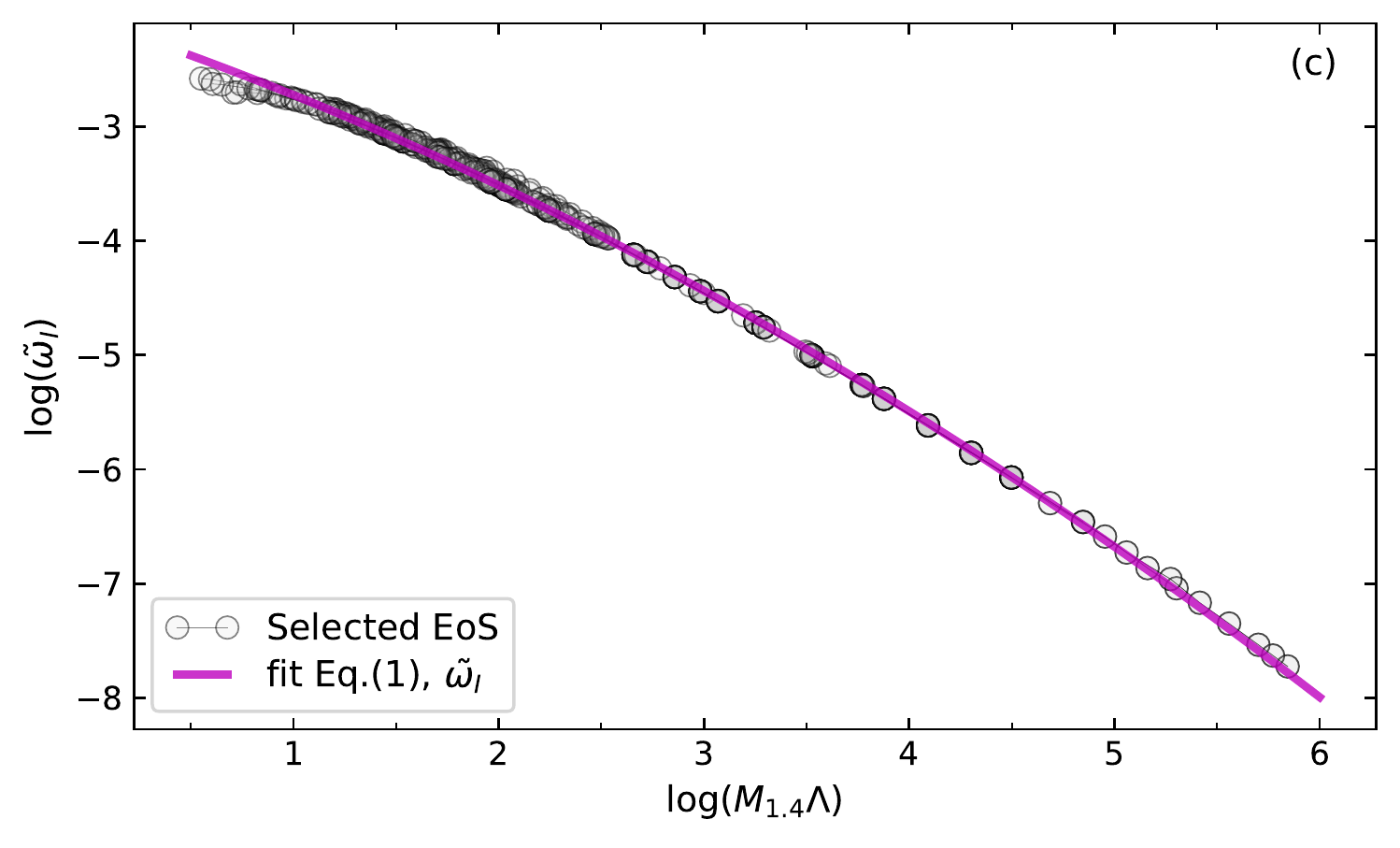}
\includegraphics[width=0.475\linewidth,angle=0]{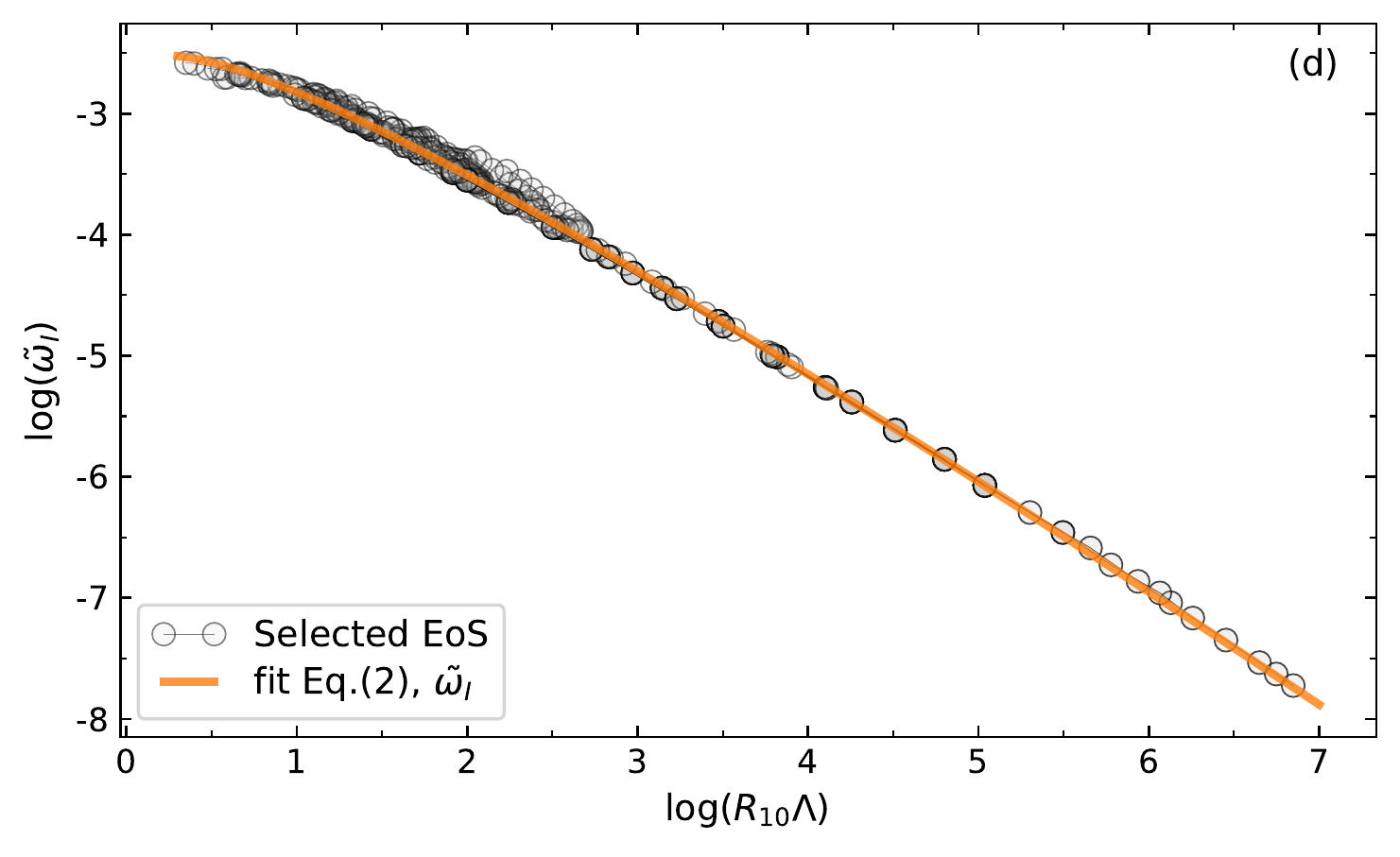}
\caption{Universal relationships for the $wI^{(0)}$ mode given by Eq.~(\ref{UR0a}) in the left panels and Eq.~(\ref{UR0b}) in the right panels. Top (bottom) panels show URs for the frequency (damping time) of the mode. {With black empty circles we present synthetic data from the selected EoS.}} 
\label{fig:URw0}
\end{figure*}

\section{Universal Relationships that include SSHS}\label{sec:newUR}

Following \citet{lugones2021arXiv} and \citet{RSetal2022PRD}, we have constructed HSs using a {\it soft} and a {\it stiff} hadronic EoS implementing the formalism proposed by \cite{OBoyle-etal-2020}. {Despite being only two, these parametric EoS are constructed in order to act as enveloping curves for a wide variety of modern hadronic EoS. {This} choice {is also supported by} the fact that the breaking of the URs comes from the slow-stable extended branch and it depends mostly on the quark EoS model.}

As we considered the hadron-quark phase transition to be sharp, we have used, for the quark-sector EoS, the constant speed of sound (CSS) parametrisation \citep{css-original}, varying its three free parameters: transition pressure, $p_t$, energy density jump, $\Delta ilon$ and speed of sound, $c_s$. This general approach allows us to study a wide range of possible scenarios in a model-independent fashion.

We have constructed a total of over $5000$ hybrid EoSs using the mentioned soft and stiff hadronic EoS, and the following parameter ranges for the CSS quark EoS:
\begin{align*}
    10~ \mathrm{MeV/fm^3} &\le p_{t} \le 300~\mathrm{MeV/fm^3} \, , \\
    100~\mathrm{MeV/fm^3} &\le \Delta \epsilon \le 3000~\mathrm{MeV/fm^3} \, , \\
    0.2 &\le c_\mathrm{s}^2 \le 1 \, .
\end{align*}
We have selected twenty of them, that represent the qualitative behaviour of all the EoSs that we have constructed, verify \mbox{$2M_{\odot} < M_{\max}  < 2.3M_{\odot}$} \citep{Rezzolla_2018,Shibata:2019ctb} and are consistent with modern astronomical observations from GW170817 \citep{GW170817-detection} and the NICER Collaboration \citep{Riley2019,Miller2019,riley2021ApJ-j0740,miller2021ApJ-j0740}, shown in Fig.~\ref{fig:MR}. {This selection {allows} us to present a wide variety of cases that capture, in a model-independent manner, the behaviour of matter inside compact objects. }

\begin{table}
\centering
\begin{tabular}{ccc} 
\toprule
\multicolumn{3}{c}{Universal relationships}\\
\multicolumn{3}{c}{$\log(\tilde{\omega}_{R,I}) = a_{R,I} + b_{R,I} x + c_{R,I} x^2$} \\ \midrule
& value $(\pm)$ & error ($\%$) \\ \cmidrule{2-3}
$a_R$ & $1.052 \;(\pm 0.0022)$ & $0.21$ \\ \midrule
$b_R$ & $-0.984 \;(\pm 0.0017)$ & $0.17$ \\ \midrule
$c_R$ & $-0.030 \;(\pm  0.0003)$ & $0.99$ \\ \midrule
$a_I$ & $-2.065 \;(\pm 0.0243)$ & ${1.16}$ \\ \midrule
$b_I$ & $-0.594 \;(\pm 0.0061)$ & $1.01$ \\ \midrule
$c_I$ & $-0.066 \;(\pm 0.0011)$ & $1.51$ \\ \bottomrule
\end{tabular}
\caption{Best values and corresponding errors for the fits proposed in Eq. \eqref{UR0a}.
 The sum of squares of residuals is  0.044 (0.601).}
\label{tab:param-univ-rel-0}
\end{table}

\begin{table}
\centering
\begin{tabular}{ccc}
\toprule
\multicolumn{3}{c}{Universal relationships}\\
\multicolumn{3}{c}{$\log(\tilde{\omega}_{R,I}) = \alpha_{R,I} + \beta_{R,I} y + \gamma_{R,I}\sqrt{y}$} \\ \midrule
 & value $(\pm )$ & error ($\%$) \\ \cmidrule{2-3}
$\alpha_R$ & $0.885 \;(\pm 0.0024)$ & $0.27$ \\ \midrule
$\beta_R$ & $-0.965 \;(\pm 0.0008)$ & $0.08$ \\ \midrule
$\gamma_R$ & ----- & ----- \\ \midrule
$\alpha_I$ & $-2.794 \;(\pm 0.0245 )$ & $0.88$ \\ \midrule
$\beta_I$ & $-1.151 \;(\pm 0.0092 )$ & $0.80$ \\ \midrule
$\gamma_I$ & $1.123 \;(\pm 0.0306 )$ & $2.72$\\ \bottomrule
\end{tabular}
\caption{Best values and corresponding errors for the fits proposed in Eq. \eqref{UR0b}. 
The sum of the squares of the residuals is 0.340 (0.909).}
 \label{tab:param-univ-rel-1}
\end{table}

{For all these {EoSs}, we have solved the TOV equations that determine the stellar structure of compact objects. Moreover, we have taken into account the impact of considering the effect of the hadron-quark phase transition on the dynamical stability of HSs and have considered the slow conversion scenario as a working hypothesis\footnote{{It is important to remark that the results of considering the rapid conversion scenario are included in the ones presented in this work, just having in mind that under this assumption, stable stellar configurations are only those up to the maximum mass of each curve in the mass-radius diagram presented in Fig. \ref{fig:MR}.}}. Regardless of this hypothesis, stable branches of stellar configurations are separated from unstable ones at the configuration for which the fundamental radial eigenfrequency is zero, as stellar stability definition indicates. Such configuration is called \emph{terminal mass} and, in the case of slow hadron-quark conversion, it does not, in principle, coincide with the critical points of the mass-radius diagram \citep[for a more detailed description see, for example,][]{Pereira:2017rmp,LugGrunf-universe:2021,RSetal2022PRD}, as it does for pure hadronic NSs of for HSs with rapid conversions.}

{To obtain more general and representative results in order to propose new URs, we have included in our analysis hybrid EoSs that produce long extended branches of SSHSs as well as long branches of {\it totally} stable HSs. In this case, looking at Fig.~\ref{fig:MR}, the SSHSs correspond to branches after the maximum mass configuration where the mass decreases as a function of the central energy density, $\epsilon_c$, and are only stable within the slow conversion hypothesis; otherwise, the totally stable HSs correspond to hybrid branches where the mass increases with the central energy density  \mbox{-e.g., traditional twin branches-} and are stable independently of the hadron-quark speed conversion.
In this sense, it is worth noticing that the results and figures presented in this work display, considering our working hypothesis, only stable stellar configurations; e.g., all the points in the mass-radius relationships, given by grey curves in Fig.~\ref{fig:MR}, are stable stellar configurations and the terminal mass is given by the last point of each curve.}

\begin{figure*}
\centering
\includegraphics[width=0.475\linewidth,angle=0]{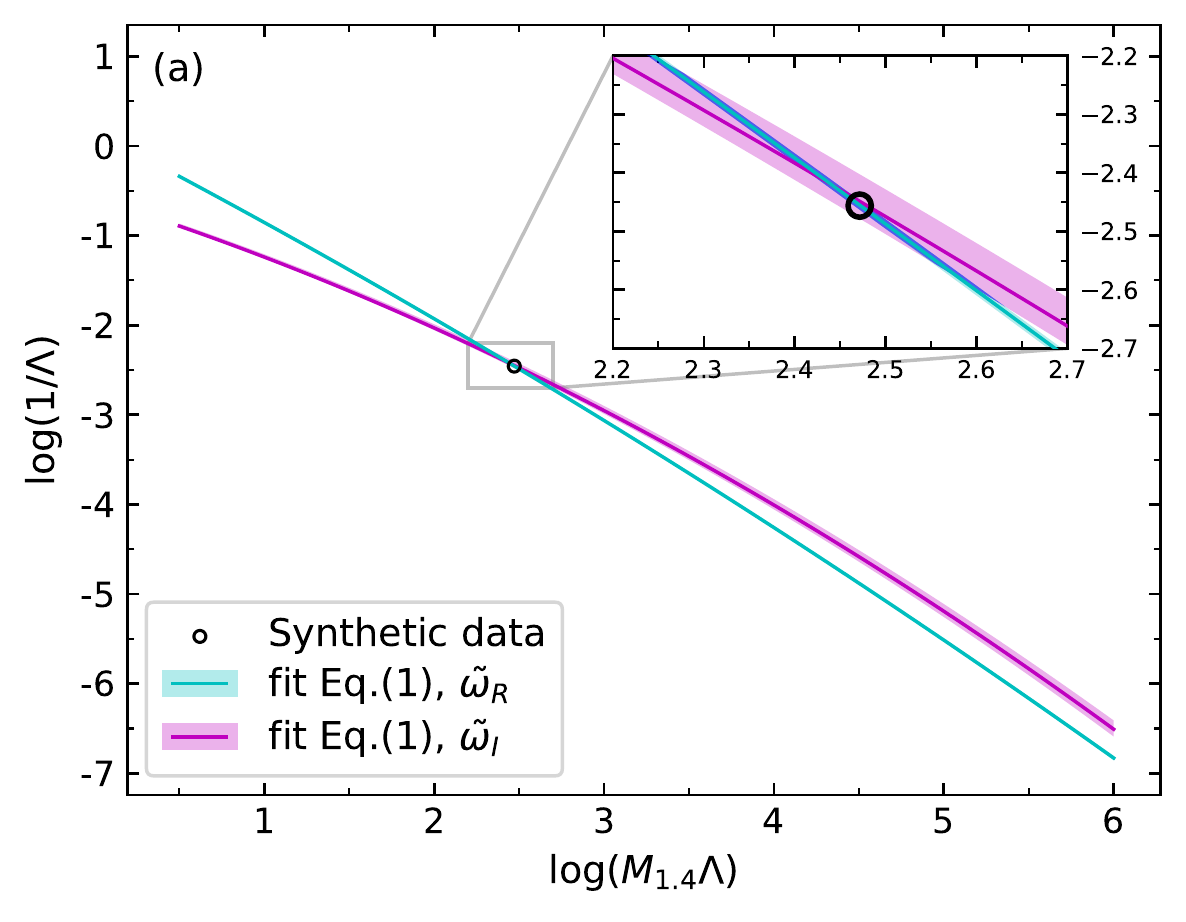}
\includegraphics[width=0.475\linewidth,angle=0]{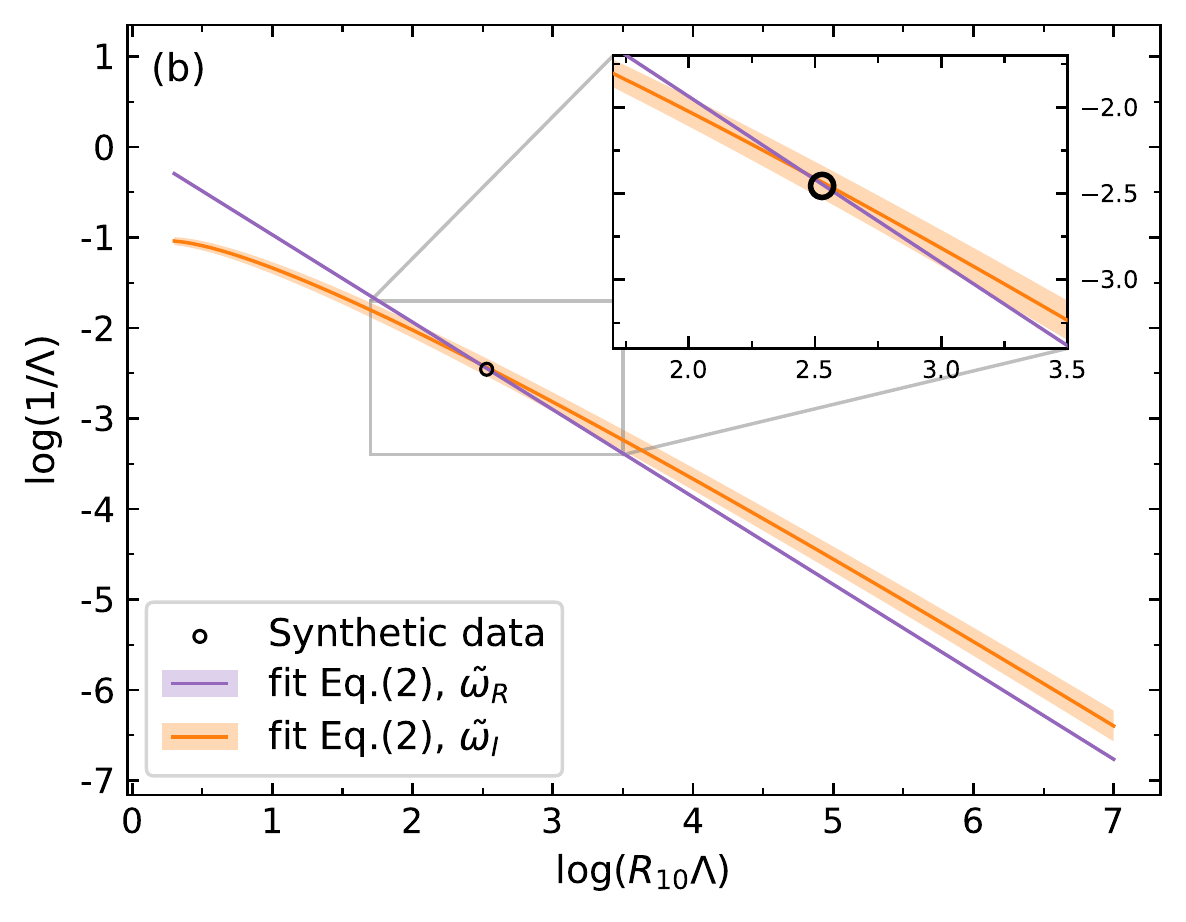}
 \caption{Determination of $M$, $R$ and $\Lambda$ of a stellar configuration through the detection of its $wI^{(0)}$ mode, using URs (and its respective error bands) given by Eq.~(\ref{UR0a}) (panel a) and Eq.~(\ref{UR0b}) (panel b). The enlarged area shows the error bands of each UR. {We represent with} a circle {the} synthetic data of the example {corresponding to the object called NS$_{1.4}^\mathrm{soft}$ whose parameters are presented in Table \ref{tabla:application}}. The predicted values --given by the intersection of the fit curves-- are in great agreement with the synthetic data --given by the black circled dot--.} 
\label{fig:cruces}
\end{figure*}

As shown in \citet{RSetal2022PRD}, SSHS break the URs for axial $wI$ modes previously proposed in the literature. Thus, using the selected hybrid EoSs, we have calculated both the fundamental and first overtone quadrupolar $wI$ axial modes using the {\it{Frequency Identificator Library}} (FIdeL) developed in \citet{RSetal2022PRD} and 
%
%
we have obtained two different URs valid for the fundamental quadurpolar axial mode, $wI^{(0)}$.  Defining $\tilde{\omega}_R = \nu / \Lambda$ and $\tilde{\omega}_I = 1/(\Lambda \tau)$, we found that $\log (\tilde{\omega}_R)$ and $\log (\tilde{\omega}_I)$ follow, for all our models, simple tight correlations with $x=\log (M_{1.4}/\Lambda)$ and $y=\log (R_{10}/\Lambda)$, were $M_{1.4}=M/1.4M_\odot$ and $R_{10}=R/10$km (see Fig.~\ref{fig:URw0}). The best fits obtained follow simple functional relationships:
\begin{equation}
\log(\tilde{\omega}_{R,I}) = a_{R,I} + b_{R,I} x + c_{R,I}x^2 \,, \label{UR0a}
\end{equation}
and
\begin{equation}
\log(\tilde{\omega}_{R,I}) = \alpha_{R,I} + \beta_{R,I} y + \gamma_{R,I}\sqrt{y} \,, \label{UR0b}
\end{equation}
where the coefficients of the best fits we have found are shown in Table~\ref{tab:param-univ-rel-0} and Table~\ref{tab:param-univ-rel-1}, respectively. In order to obtain those numerical values, the units for the frequency and damping time must be kHz and $\mu$sec, respectively.

We present, for the sake of completeness, the URs for $\log(\tilde{\omega}_{R,I})$ as a function of $ z=\log(\Lambda M_{1.4}/R_{10})$, valid for the first overtone, $wI^{(1)}$. The best fit relationships we were able to find read
 \begin{equation}
     \log(\tilde{\omega}_{R,I}) = \delta_{R,I} + \kappa_{R,I} z + \eta_{R,I}z^2 \,. \label{UR1}
 \end{equation}
{Since in the following section we do not present an asteroseismology application for the URs of the first overtone, and in order to keep clarity considering the main scope of our work, we present the details of the URs for $wI^{(1)}$ in Appendix~\ref{app:fits}; there,} coefficients, errors and fitting test of the best fits are presented in Table~\ref{tab:param-univ-rel-w1}. To see the tightness of the fits, see Fig. \ref{fig:URw1}.

All the relationships that we present have some degree of degeneracy, {since the curves for HSs and SSHSs overlap with part of the curves for purely hadronic NSs.} This can be seen more clearly in panel (d) of Fig.~\ref{fig:URw0}, where the tight cloud of synthetic data presents a larger dispersion. However, as we will show in the next section, this intrinsic degeneracy is broken by the asteroseismological method proposed below.

\section{Asteroseismology of axial \textit{wI} modes revisited} \label{sec:astro}



If the $wI^{(0)}$ mode is detected, and both its frequency{, $\nu_0$,} and damping time{, $\tau_0$,} are measured, the URs presented in Eq.~(\ref{UR0a}) can be used to determine $M$ and $\Lambda$ of the pulsating object, and the URs presented in Eq.~(\ref{UR0b}) allow to determine $R$ and $\Lambda$. 
The general idea is that if $\nu_0$ and $\tau_0$  of the fundamental axial $wI$ mode are known, the URs given in Eq.~(\ref{UR0a}) allow, after rearranging the $\log(\nu _0)$ and $\log(\tau_0)$ terms, to isolate $\log(1/\Lambda)$ {on} the left {hand} side of Eq.~(\ref{UR0a}). This procedure provides two different curves in the $\log(1/\Lambda)$-$\log(M_{1.4}\Lambda)$ plane, corresponding to the $R$ and $I$ parts of this equation. Equivalently, Eq.~(\ref{UR0b}) becomes a pair of curves in the  $\log(1/\Lambda)$-$\log(R_{10}\Lambda)$ plane. The intersection of these curves in each of the two planes allows estimating $M$, $R$, and $\Lambda$ of the pulsating object{. In Fig.~\ref{fig:cruces}, we present an illustration of the method applied to a specific example corresponding to the object called NS$_{1.4}^\mathrm{soft}$ whose parameters are presented in Table~\ref{tabla:application}. In this figure, given URs proposed in Eqs. (\ref{UR0a}) and (\ref{UR0b}) and the supposed observed values of $\nu_0$ and $\tau_0$ for this object, we present the resulting prediction of the asteroseismology method due to the curve intersections in each plane. There, it can be seen graphically the potential precision of the method when comparing with the synthetic data, given in the figure by the black circles.}

\begin{table*}
\centering
\begin{tabular}{ccccccccccc}
\toprule
 & \multicolumn{5}{c}{Synthetic data} & \multicolumn{3}{c}{URs of Eqs.~\eqref{UR0a} and \eqref{UR0b}} & \multicolumn{2}{c}{URs of \citet{Benhar2004PRD}} \\
 \cmidrule(r){2-6} \cmidrule(lr){7-9} \cmidrule(l){10-11}
 & $M~[M_\odot]$ & $R~[\mathrm{km}]$ & $\Lambda$  & $\nu~[\mathrm{kHz}]$ & $\tau~[\mu\mathrm{sec}]$ & $M~[M_\odot]$ & $R~[\mathrm{km}]$ & $\Lambda$ & $M~[M_\odot]$ & $R~[\mathrm{km}]$ \\
\midrule
NS$_{1.4}^\mathrm{soft}$ & $1.45$ & $11.8$ & $285.8$ & $7.78$ & $30.56$ & $1.47^{\nfrac{+0.111}{-0.099}}$ & $12.0^{\nfrac{+0.9}{-0.5}}$ & $263^{\nfrac{+273}{-114}}$ &  $1.4^{\nfrac{+0.42}{-0.33}}$ & $12^{\nfrac{+1.4 }{-2.4}}$ \\ \midrule
HS$_{1.4}^\mathrm{soft}$ & $1.43$ & $11.0$ & $170.1$ & $8.39$ & $30.60$ & $1.42^{\nfrac{+0.083}{-0.095}}$ & $11.0^{\nfrac{+0.7}{-0.5}}$ & $192^{\nfrac{+172}{-78}}$ &  $1.4^{\nfrac{+0.40}{-0.32}}$ & $11^{\nfrac{+1.4 }{-2.3}}$ \\ \midrule
SSHS$_{1.4}^\mathrm{soft}$ & $1.44$ & $10.2$ & $67.4$ & $9.13$ & $32.87$ & $1.41^{\nfrac{+0.066}{-0.072}}$ & $9.8^{\nfrac{+0.5}{-0.3}}$ & $100^{\nfrac{+70}{-35}}$ &  $-$ & $-$ \\ \midrule
NS$_{1.4}^\mathrm{stiff}$ & $1.44$ & $13.4$ & $696.9$ & $7.00$ & $29.78$ & $1.49^{\nfrac{+0.125}{-0.141}}$ & $13.7^{\nfrac{+1.3}{-0.7}}$ & $489^{\nfrac{+694}{-238}}$ &  $1.4^{\nfrac{+0.43}{-0.33}}$ & $14^{\nfrac{+1.4}{-2.3}}$ \\ \midrule
SSHS$_{1.4}^\mathrm{stiff}$ & $1.45$ & $12.2$ & $261.7$ & $7.77$ & $30.13$ & $1.45^{\nfrac{+0.106}{-0.101}}$ & $12.0^{\nfrac{+0.9}{-0.6}}$ & $285^{\nfrac{+306}{-98}}$ &  $-$ & $-$ \\ \midrule
NS$_{2}^\mathrm{soft}$ &  $2.05$ & $11.7$ & $23.37$ & $6.94$ & $57.62$ & $2.04^{\nfrac{+0.062}{-0.069}}$ & $12.2^{\nfrac{+0.6}{-0.5}}$ & $23^{\nfrac{+11}{-6}}$ &  $2.1^{\nfrac{+0.49}{-0.42}}$ & $12^{\nfrac{+0.7 }{-3.2}}$  \\ \midrule
SSHS$_{2}^\mathrm{soft}$ & $2.00$ & $11.1$ & $16.3$ & $7.28$ & $61.07$ & $2.00^{\nfrac{+0.051}{-0.058}}$ & $11.6^{\nfrac{+0.4}{-0.5}}$ & $16^{\nfrac{+7}{-4}}$ &  $-$ & $-$ \\ \midrule
NS$_2^\mathrm{stiff}$ & $2.07$ & $13.6$ & $73.2$ & $6.30$ & $48.40$ & $2.05^{\nfrac{+0.093}{-0.104}}$ & $14.1^{\nfrac{+0.7}{-0.5}}$ & $65^{\nfrac{+43}{-23}}$ &  $2.1^{\nfrac{+0.55}{-0.45}}$ & $14^{\nfrac{+2.1 }{-3.4}}$ \\ \midrule
HS$_2^\mathrm{stiff}$ & $2.18$ & $13.0$ & $32.1$ & $6.41$ & $57.10$ & $2.16^{\nfrac{+0.069}{-0.088}}$ & $13.5^{\nfrac{+0.6}{-0.5}}$ & $31^{\nfrac{+15}{-10}}$ & $2.2^{\nfrac{+0.54}{-0.46}}$ & $13^{\nfrac{+2.2 }{-3.4}}$ \\ \midrule
SSHS$_2^\mathrm{stiff}$ & $2.01$ & $12.1$ & $26.3$ & $7.09$ & $54.17$ & $1.98^{\nfrac{+0.060}{-0.074}}$ & $12.2^{\nfrac{+0.5}{-0.3}}$ & $28^{\nfrac{+14}{-8}}$  & $-$ & $-$ \\ 
\bottomrule
\end{tabular}
\caption{Results of applying the proposed URs given by Eqs.~(\ref{UR0a}) and (\ref{UR0b}) to ten synthetic objects. For each case, we show $M$, $R$, $\Lambda$ obtained after solving the TOV equations and $\nu$, $\tau$ of the fundamental axial $wI$ mode. The prediction and associated error bands of applying the proposed URs  for such astronomical quantities is shown. Except for SSHSs, estimates and error bands for the mass and radius obtained using the relationships from \citet{Benhar2004PRD} are also presented.}
\label{tabla:application}
\end{table*}

\begin{figure}
\centering
\includegraphics[width=0.95\linewidth,angle=0]{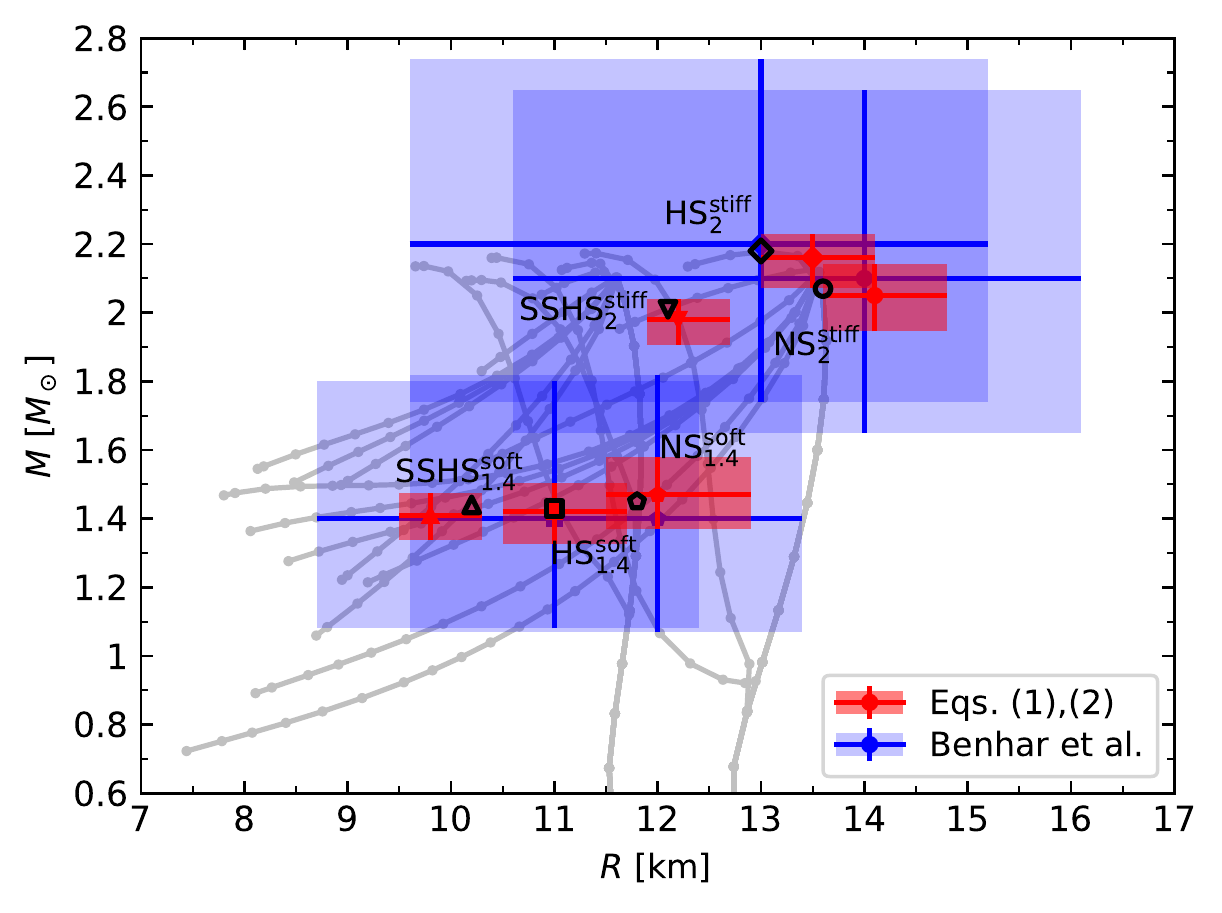}
\caption{{Mass-Radius predictions and their associated errors}.  With black symbols, we display some synthetic objects of Table~\ref{tabla:application}, specifically the $1.4~M_\odot$~soft and the $2~M_\odot$~stiff cases, as an example of the results given by the asteroseismology method. The red dots -and the associated error bars and coloured rectangle boxes- are the associated predictions using the URs proposed in Eqs.~(\ref{UR0a}) and (\ref{UR0b}). The blue dots, bars, and coloured boxes correspond to the predictions using the UR proposed by \citet{Benhar2004PRD}. Although both predictions lie reasonably near the synthetic data, the error boxes of the URs proposed in our work are significantly smaller than the ones of \citet{Benhar2004PRD}. Since the \citet{Benhar2004PRD} URs do not consider SSHSs we do not include their predictions for these objects. } 
\label{fig:mrboxes}
\end{figure}

We present ten gedankenexperiments to show the capabilities of this method to obtain estimates for $M$, $R$, and $\Lambda$. In addition, we compare our results with those obtained using the relationships from \citet{Benhar2004PRD}. The examples include a wide range of different compact objects that can be separated into two different groups.
The first one contains $1.4\,M_\odot$ compact objects: NS$_{1.4}^\mathrm{soft}$ and NS$_{1.4}^\mathrm{stiff}$ (purely hadronic NSs constructed with the {soft} and {stiff} hadronic EoS, respectively), HS$_{1.4}^\mathrm{soft}$ (a {totally} stable HS), SSHS$_{1.4}^\mathrm{soft}$ and SSHS$_{1.4}^\mathrm{stiff}$ (SSHSs constructed with hybrid EoS but with the {soft} and {stiff} hadronic one, respectively). The second group contains high-mass compact objects with masses around $2\,M_\odot$: NS$_2^\mathrm{soft}$ and NS$_2^\mathrm{stiff}$ (purely hadronic NS constructed with the {\it soft} and {\it stiff} hadronic EoS, respectively), HS$_2^\mathrm{stiff}$ (a totally stable HS), SSHS$_2^\mathrm{soft}$ and SSHS$_2^\mathrm{stiff}$ (SSHSs  constructed with hybrid EoS with the {soft} and {stiff} hadronic one, respectively).

In Table~\ref{tabla:application}, we present the synthetic data ($M$, $R$, $\Lambda$ and $\nu$, $\tau$ of the fundamental $wI$ mode) for the ten selected cases and the predicted values (with their corresponding errors) obtained from our proposed URs.  For comparison, we also show the predictions of \citet{Benhar2004PRD} except for the SSHS cases where they are not valid. The predictions of $M$, $R$, and $\Lambda$ for each synthetic object are obtained using the corresponding fitting coefficients  given in Tables~\ref{tab:param-univ-rel-0} and \ref{tab:param-univ-rel-1} and applying the procedure described at the beginning of this section. The errors in $M$, $R$, and $\Lambda$ are derived from the crossings of the envelope curves constructed with the error coefficients of Tables~\ref{tab:param-univ-rel-0} and \ref{tab:param-univ-rel-1}.
The envelope curves between the four crossing points define an area. The errors are derived from the projection of this area on the axes $\log(1/\Lambda)$, $\log(M_{1.4}\Lambda)$ and $\log(R_{10}\Lambda)$.
For the dimensionless tidal deformability, $\Lambda$, two independent estimates are available, one from the URs given in Eq.~(\ref{UR0a}) and the other from Eq.~(\ref{UR0b}). In all the studied objects both estimates are consistent with each other and the one obtained with Eq.~(\ref{UR0a}) is completely enclosed within the estimate of Eq.~(\ref{UR0b}). {For this reason, we use the estimate obtained with Eq.~(\ref{UR0a}).}

In Fig.~\ref{fig:mrboxes}, we present the estimate and error boxes for some of the selected cases, specifically the $1.4~M_\odot$~soft and the $2~M_\odot$~stiff cases. We find that the proposed URs are capable of distinguishing hadronic NSs from HSs and SSHSs both for 1.4 $M_\odot$ and 2 $M_\odot$ objects. Moreover, a visual comparison with the error boxes obtained using the relationships proposed by \citet{Benhar2004PRD} {shows that our URs significantly improve the predicted values of $M$ and $R$. These results suggest that asteroseismology with $wI$ modes
{could help test} the SSHS hypothesis and may give theoretical insight on the nature of the hadron-quark phase transition.}
%

\begin{figure}
\centering
\includegraphics[width=0.95\linewidth,angle=0]{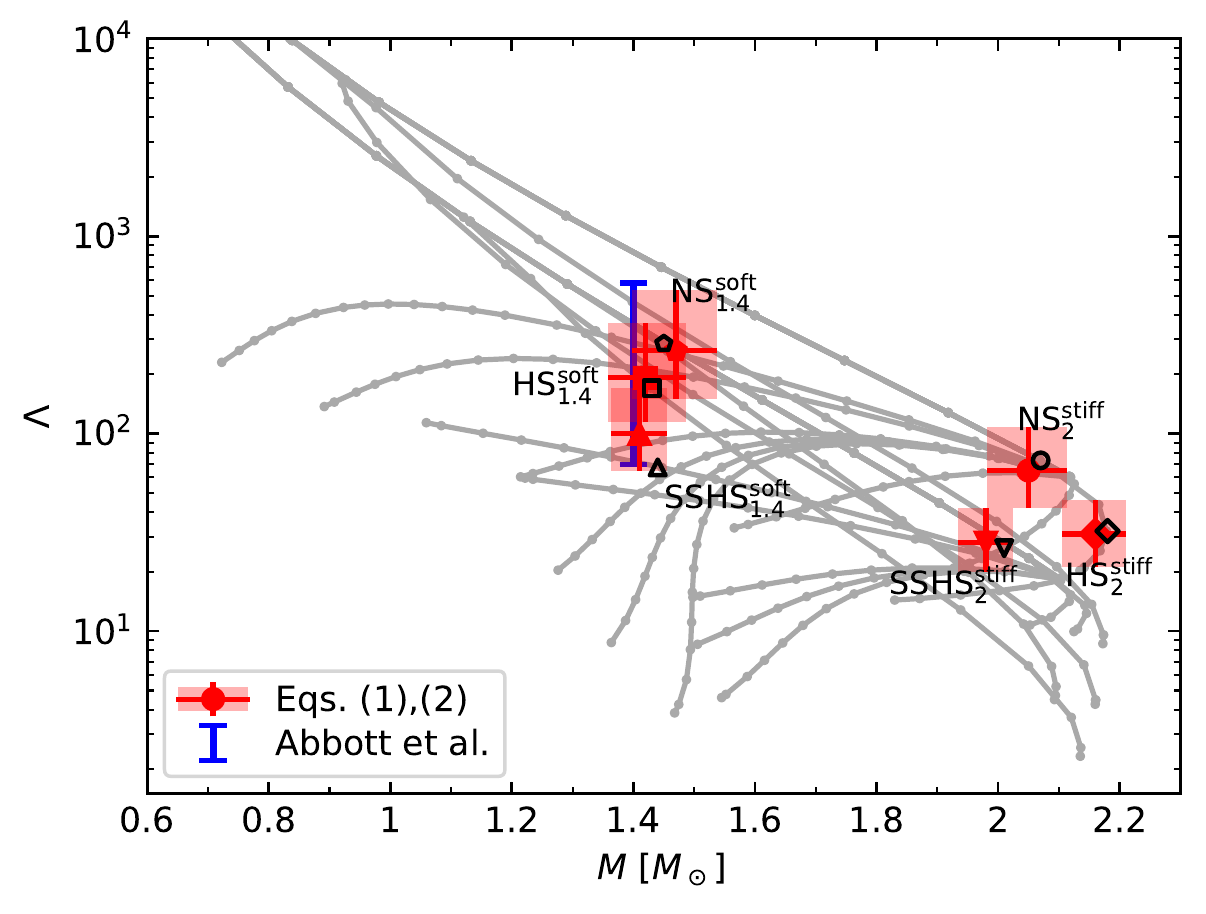}
\caption{{Tidal deformability predictions and their associated errors.} Black symbols represent the $1.4~M_\odot$~soft and the $2~M_\odot$~stiff synthetic objects of Table~\ref{tabla:application}.   The red dots -and the associated error bars and coloured rectangle boxes- are the predictions coming from Eq. \eqref{UR0a}. The blue error bar indicates the constraint imposed by the GW170817 event \citep{Abbott:2018}.}
\label{fig:tidalboxes}
\end{figure}

In Fig.~\ref{fig:tidalboxes}, we present the predictions for the mass and the dimensionless tidal deformability of some selected objects of Table~\ref{tabla:application}. 
The error boxes coming from Eq. \eqref{UR0a} {appear to be} smaller than the observational error bar of the GW170817 event and its electromagnetic counterpart \citep{Abbott:2018}. {This suggests that in principle} the here-proposed asteroseismology method could {help in the future to} place significant constraints on $\Lambda$.

\section{Summary and Conclusions} \label{sec:conc}

Motivated by the results presented by \citet{RSetal2022PRD}, {where} we showed  {that axial $wI$ mode URs in the literature fail for SSHSs}, we have presented {new} URs that involve (like \citet{sotanikumar2021} for polar modes) not only $M$ and $R$ but also $\Lambda$ for the two first axial $wI$ modes of compact stars. These new versions are valid for SSHSs, a feature not included in any of the URs available so far.
 
We have shown the capabilities of these new URs to determine, after a detection of both the frequency and damping time of $wI^{(0)}$, the value of $M$, $R$, and $\Lambda$ of the pulsating compact object with small errors. {Moreover, w}e have shown that the errors of the estimates that our URs can produce are better than those of \citet{Benhar2004PRD} even for {totally} stable compact objects where that relationship is valid. The analysed  examples indicate that the estimate of $M$ is improved by approximately an order of magnitude and that the error for $R$ reduces, approximately, by more than a half. Moreover, on average, our estimate of $\Lambda$ involves an error of $\sim 50\%$. If we concentrate only on objects of $M\sim 1.4\, M_\odot$, the average error is $\sim 70\%$. Although this estimation could seem {better} than the \mbox{$\Lambda_{1.4} = 190^{\scriptscriptstyle +390}_{\scriptscriptstyle -120}$} estimate obtained from GW170817 and its electromagnetic counterpart \citep{Abbott:2018}, it is important to remark that this comparison is made between present time estimations and (potential) future detection of such axial modes. This is not a completely fair comparison as estimations of $\Lambda$ coming from binary NS mergers, by the time $wI$ modes become detectable, should have improved consistently.

The URs presented in this work could be further improved, with a small change in the fitting parameters, if the slow conversion scenario is refuted and SSHSs are not possible. In such a case, a new fit of Eqs.~(\ref{UR0a}) and (\ref{UR0b}) excluding slow-stable objects would render even better estimates  of $M$, $R$, and $\Lambda$. 

{The proposed method could eventually help discriminate between purely hadronic, totally stable HSs, and SSHSs. In the best case scenario, multiple detections of $wI^{(0)}$ modes of different stars with similar mass -e.g., a detection of a NS$_2^\mathrm{stiff}$ type star and of a SSHS$_2^\mathrm{stiff}$ one- could be interpreted as strong evidence of the existence of multiple stable branches of compact objects and shed some light {on} the validity of the SSHS hypothesis\footnote{{Similar conclusions might be possible with detection of $wI^{(0)}$ modes of a NS$_{1.4}$ and SSHS$_{1.4}$.}}. In any case, the method might become a useful tool to understand, together with $M$ and $R$ estimates obtained from electromagnetic observations \citep[with instruments like NICER ][]{Miller2019,Riley2019,miller2021ApJ-j0740,riley2021ApJ-j0740}, the behaviour of matter under extreme conditions.}

{An alternative (and complementary) way of testing the SSHS hypothesis is through the detection of fluid modes. Particularly interesting is the fact that the frequency of the quadrupolar fundamental mode for SSHSs becomes significantly higher when compared to a NS with the same mass \citep[see, for example, ][]{Tonetto:2020bie,mariani2022MNRAS}. Moreover, a distinctive feature that pinpoints the occurrence of a slow sharp hadron-quark phase transition in the inner core of HSs is the appearance of a $g$ mode associated with such phase transition \citep[see, for example, ][]{Tonetto:2020bie,Rodriguez-etalg2,mariani2022MNRAS}\footnote{{$g$ modes due to HSs with discontinuous hybrid EoS have also been calculated in \citet{g-mode_01,g-mode_02,g-mode_03,RSetalJCAP} but without taking into account the speed of the hadron-quark conversion.}}. }


{The most promising scenarios for detecting GWs from isolated compact objects are: the post-merger phase of a binary NSs merger in which the newly born proto-neutron star is expected to emit large amounts of energy channeled into particular quasi-normal modes \citep{morozova:2018tgw,torres-forne:2019tao,Radice:2019ctg,powell:2019gwe}; the transient bursts of GWs associated with extremely energetic events in magnetars such as giant flares and pulsar (anti-)glitches \citep{corsi:2011mge,zink:2012agw,macquet:2021sfl} and the continuous semi-periodic GW-emission of rotating isolated compact objects which are asymmetric with respect to its rotational axis \citep{riles:2017rsf,sieniawska:2019cgw,cieslar:2021doc}}.

{{Regardless of these specific scenarios, t}he feasibility of detecting axial $wI$ modes is not clear{. The main reason for this is} the lack of modern numerical simulations that allow us to understand to what extent they can be excited {in astrophysical scenarios. Despite this, it has been argued by pioneering numerical calculations performed by \citet{andersson:1996gwa} and \citet{allen:1998gwf} that $w$ modes should be important during the formation of a NSs as a consequence of an asymmetrical collapse or a binary NS merger. The question related to whether such modes can be excited beyond the detection threshold of the next-generation gravitational-wave observatories like the Einstein Telescope \citep[{see, for example, }][{and references therein}]{Punturo:2010zz,maggiore:2020scf} or the Cosmic Explorer \citep[{see, for example, }][{and references therein}]{Hall:2021gwp} remains an open issue that needs to be further investigated.}} {Compared to those of a-LIGO, the sensitivity of third-generation detectors is expected to improve by at least an order of magnitude in the frequency range associated with $wI^{(0)}$ modes. Due to this fact, the possibilities of detecting GWs from isolated compact objects (and particularly $w$ modes) are expected to get enhanced}. If axial $wI$ modes are ever observed, our URs provide an improved method to estimate $M$, $R$, and $\Lambda$ of the emitting object.

\section*{Acknowledgements}
{The authors thank the anonymous referee for the constructive
comments and criticisms that have contributed to improve the
manuscript.} I.F.R.-S.,  M.M. and O.M.G. thank CONICET and UNLP for financial support under grants G157, G007, X824. I.F.R.-S. is also partially supported by PICT 2019-0366 from ANPCyT (Argentina) and by the National Science Foundation (USA) under Grant PHY-2012152. M.M is a postdoctoral fellow of CONICET.
G.L. acknowledges the support of the Brazilian agencies Conselho Nacional de Desenvolvimento Cient\'{\i}fico e Tecnol\'ogico (grant 316844/2021-7) and Funda{\c c}\~ao de Amparo \`a
Pesquisa do Estado de S\~ao Paulo  (grants 2022/02341-9 and 2013/10559-5)

\section*{Data Availability}

The computed data presented and discussed in this paper will be shared upon reasonable request.




\bibliographystyle{mnras}
\bibliography{ifrs-bib} 


\appendix

\section{First overtone universal relationships} \label{app:fits}

\begin{figure}
\centering
\includegraphics[width=0.95\linewidth,angle=0]{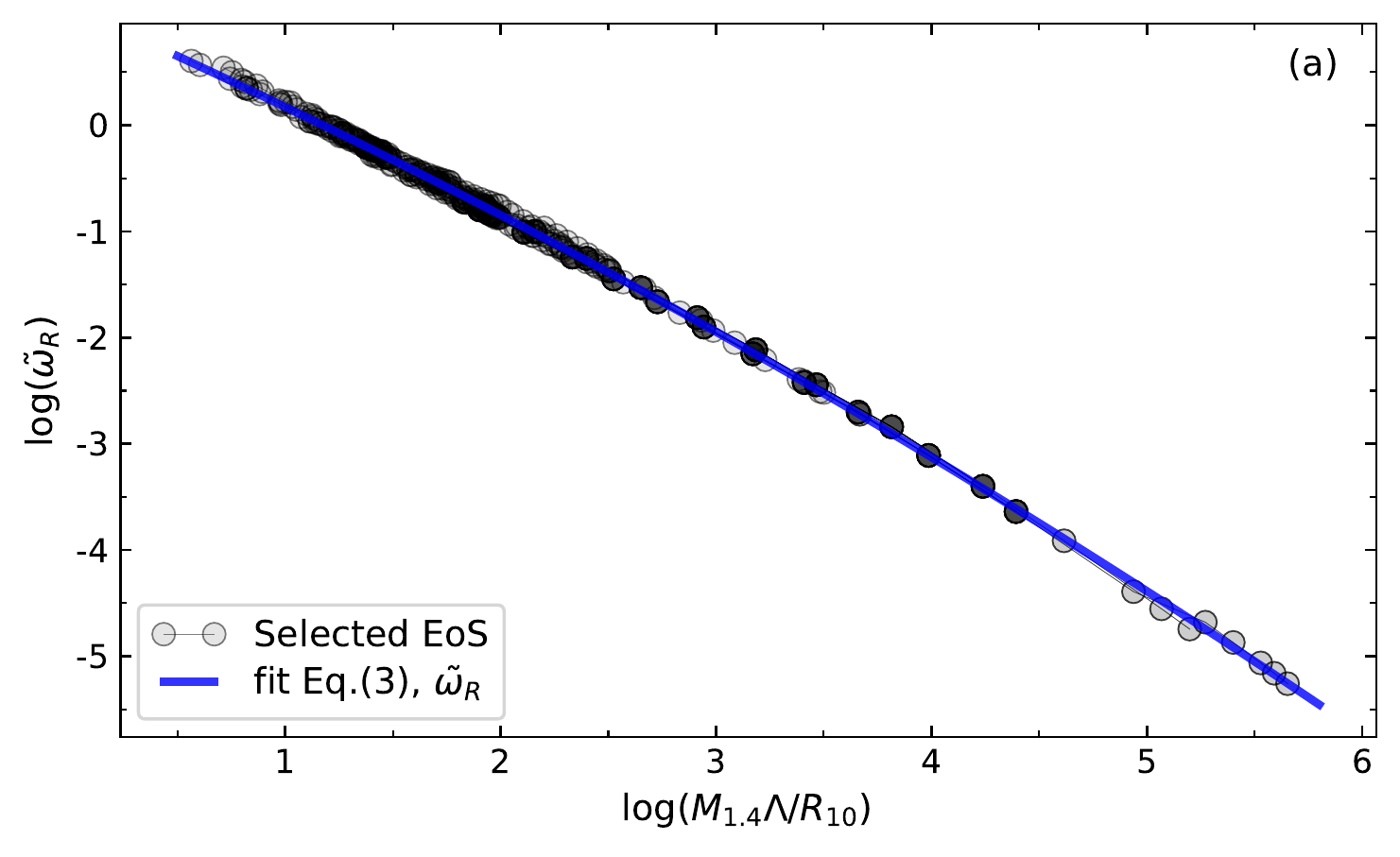}
\includegraphics[width=0.95\linewidth,angle=0]{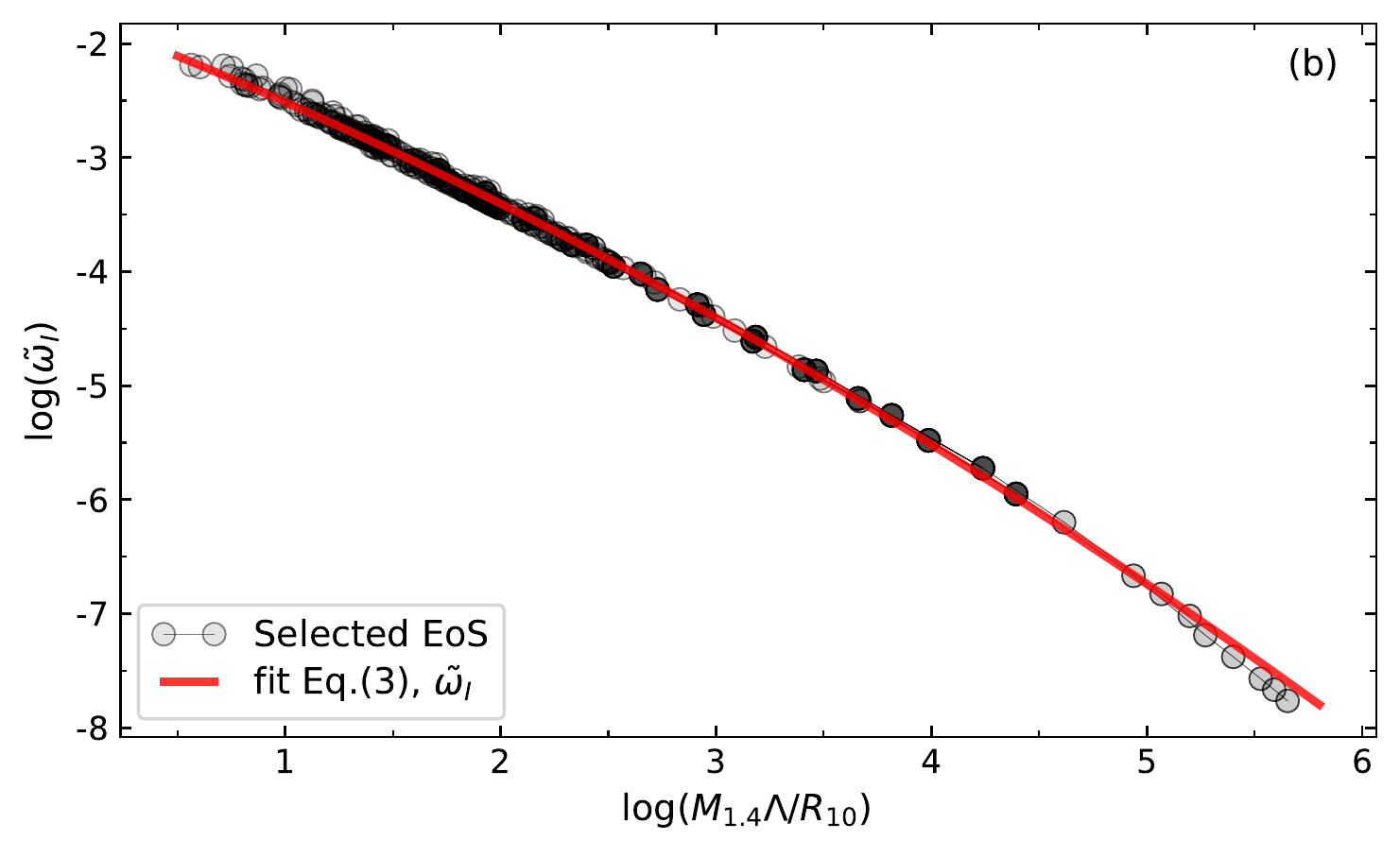}
\caption{Universal relationships for the $wI^{(1)}$ mode given by Eq.~(\ref{UR1}). Both relationships include dimensionless tidal deformability, mass, and radius. The top (bottom) panel shows the UR for the frequency (damping time) of the mode.} 
\label{fig:URw1}
\end{figure}

In this appendix, we present the URs for the first overtone, $wI^{(1)}$. The coefficient values, errors, and fitting test for the URs of Eq.~(\ref{UR1}) are presented in Table~\ref{tab:param-univ-rel-w1}. In order to obtain these numerical values, the units for the frequency and damping time must be kHz and $\mu$sec, respectively. Moreover, in Fig.~\ref{fig:URw1}, we present the URs for the first overtone from which the tightness of the proposed fit can be clearly seen. 

{{If our URs were applied, the detection of a $wI^{(1)}$ mode would not allow to determine $M$, $R$ and $\Lambda$ but only $\Lambda$ and $M/R$. Moreover, the relationships presented for the first overtone in previous works, such as \citet{Tsui2005MNRAS,w-modes_universal}, do not provide error bars that allow the method comparisons. For this reason, we are not including any astrophysical asteroseismological application of these URs and we present the details of these fits in this appendix. However, as can be seen, both graphically and in the values of errors and residuals of our fits, the URs proposed are able to include different branches of stellar stability, including the SSHS; as shown in \citet{RSetal2022PRD}, the others URs do not.}}



\begin{table}
\centering
\begin{tabular}{ccc} 
\toprule
\multicolumn{3}{c}{Universal relationships}\\
\multicolumn{3}{c}{$\log(\tilde{\omega}_{R,I}) = \delta_{R,I} + \kappa_{R,I} z + \eta_{R,I} z^2$} \\ \midrule
 & value $(\pm )$ & error ($\%$) \\ \cmidrule{2-3}
$\delta_R$ & $1.110 \;(\pm 0.0088)$ & $0.79$ \\ \midrule
$\kappa_R$ & $-0.896 \;(\pm 0.0067)$ & $0.75$ \\ \midrule
$\eta_R$ & $-0.0406 \;(\pm  0.0011)$ & $2.71$ \\ \midrule
$\delta_I$ & $-1.733 \;(\pm 0.0106)$ & $0.61$ \\ \midrule
$\kappa_I$ & $-0.725 \;(\pm 0.0081)$ & $1.12$ \\ \midrule
$\eta_I$ & $-0.0552 \; (\pm 0.0014)$ & $2.54$ \\ \bottomrule
\end{tabular}
\caption{Best values and corresponding errors for the fits proposed {in Eq. \eqref{UR1}}. 
The sum of squares of residuals is  0.646 (0.936).}
\label{tab:param-univ-rel-w1}
\end{table}

\bsp	
\label{lastpage}
\end{document}